\documentclass[aps,twocolumn,pra,superscriptaddress,amsmath,showpacs,tightenlines,pdflatex,longbibliography,10pt,conference]{revtex4-1}
\usepackage{amssymb}
\usepackage{amsmath}
\usepackage{dcolumn}
\usepackage{graphicx}
\usepackage{mathrsfs}
\usepackage{appendix}
\usepackage{graphicx}
\usepackage{booktabs}
\usepackage{color}

\usepackage{url}
\usepackage[colorlinks]{hyperref}
\hypersetup{%
    plainpages=true,
    breaklinks=true,
    hypertexnames=false,
    pageanchor=true,
    colorlinks=true,
    linkcolor={blue},
    citecolor={red},
    urlcolor={blue},
    anchorcolor={black}
}
\begin{document}
\title{$PT$-Symmetric magnetic Chaos in cavity magnomechanics}

\author{Mei Wang}
\affiliation{School of electrical and electronic engineering, Wuhan Polytechnic University, Wuhan, 430040, China}

\author{Duo Zhang}
\affiliation{School of electrical and electronic engineering, Wuhan Polytechnic University, Wuhan, 430040, China}

\author{Yu-Ying Wu}
\affiliation{School of electrical and electronic engineering, Wuhan Polytechnic University, Wuhan, 430040, China}

\author{Tai-Shuang Yin}
\affiliation{School of physics, Huazhong University of Science and Technology, Wuhan 430074, China}

\author{Zhao-Yu Sun}
\email{sunzhaoyu2000@qq.com}
\affiliation{School of electrical and electronic engineering, Wuhan Polytechnic University, Wuhan, 430040, China}
\date{\today}

\begin{abstract}
Here, we research a novel cavity magnomechanical system, where magnon driven by a microwave field couples with a phonon mode with a nonlinear magneostrictive interaction (radiation pressure-like).
Based on this interaction, we numerically demonstrate $PT$-symmetric chaos in this system.
With only one monochrome driving, the chaotic threshold is lowered to a rather low levels, this is due to the dynamical enhancement of nonlinearity in the $PT$-symmetry broken phase.
Moreover, by simply manipulating the phase transition between $PT$-symmetry phase and $PT$-symmetry broken phase, we can switch the system between into and out of chaotic regimes. Our work may broaden the cavity magnomechanics and provide a promising application for magnetic chaos-related security communications.
\end{abstract}

\maketitle

Very recently, the collective excitation of magnetization--magnon in magnetic materials has received considerably experimental~\cite{Chumak2015,Lenk2011,Osada2016,Cao2015, Zhang2014, Yao2015} interests which reveal the unique and attractive properties of magnon, i.e., the arbitrarily tuned frequency via manipulating the basis magnetic field~\cite{Chumak2015,Lenk2011,Serga2010} and the magnetostrictive force~\cite{Spencer1958}. This allows magnon to couple with microwave cavity and other systems to form highly tunable hybrid systems for information processing. One magnetic material that is often used in experiments is the magnetic insulator yttrium iron garnet~\cite{Serga2010} (YIG, $Y_3Fe_5O_{12}$), which has excellent material properties. It not only has rich nonlinearities but also has very low dissipation when coupling with different systems, such as: qubit, acoustic phonon, and microwave photon. Thanks to the nonlinearity, cavity magnonical~\cite{You2018} and cavity magnomehcnaical~\cite{Tang2016,Zhang2016} systems (CMMSs) are established to research many interesting quantum and classical nonlinearity phenomena, such as interacting between magnons and microwave photons in the quantum limit~\cite{Nakamura2014}, generating magnon-photon-phonon entanglement~\cite{Agarwal2018}, observing magnon-polariton bistability~\cite{You2018} and the generating magnon induced high-order sideband~\cite{Liu2018}.
Similar to the optomechanical system,
in the CMOS one strong driving is needed acting on the YIG sphere to simulate a weak (radiation pressure¨Clike) magneostrictive interaction between magnon and phonon modes. Further, the backaction-induced mechanical gain conquers the loss. Keeping increasing the driving power, the chaos motion appears in magnon and phonon modes.
Here the magneostrictive interaction in the CMMS is far below the optomechanical coupling, so the driving power and chaotic threshold are larger than that in optomechanical system.
To improve the operability of realizing magnon chaos in communication, we should decrease the driving threshold to access the chaotic regime.

In recent years, the notion of parity-time ($PT$) symmetry has been introduced into the field of optics~\cite{Bender1998,Bender2007,Peschel2012} from quantum mechanics. Based on $PT$-symmetric optical systems, extensive distinctive optical phenomena have been uncovered due to the unique physical mechanism, which are unreachable in the conventional optical systems. It works as the phase transition from the PT-symmetry phase ($PT$SP) (real eigenvalue spectrum) to spontaneously $PT$-symmetry broken phase ($PT$BP) (complex eigenvalue spectrum) by just modulating one necessary parameter in the system. The phase transition mechanism has been applied in various optical systems such as synthetic waveguides and microcavities; The corresponding representative phenomena includes experiment of optical non-reciprocity~\cite{Peng2014}, loss-induced transparency~\cite{Guo2009}, low-power optical diodes~\cite{Yang2014,Peng2014}, a single-mode laser~\cite{Feng2014} and $PT$-symmetry-breaking chaos~\cite{Lv2015}. In particular, experimentally researching on the exceptional point in cavity magnon-polaritons system~\cite{Hodaei2017,You2017} in recent has been presented. It means that the phase transition mechanics can be introduced into the hybrid magnetic systems to deepen the research of magnon features. However, the magnetic chaos in the $PT$-symmetry regime remain largely unexplored.

In our work, we explore the $PT$-symmetry and $PT$-symmetry broken magnetic chaos in a $PT$-symmetric CMMS, in which the lossy magnon mode driven by a microwave driving couples to a gain microwave cavity and a lossy phonon mode, respectively. Here we manipulate the phase transition from $PT$ symmetry to broken $PT$ symmetry to
enhance the nonlinear magnon-phonon interaction, this is in stark contrast to the normal chaos. In broken $PT$ symmetry regime, the photon localization in the active cavity results in the energy accumulation of the magnon and phonon modes with time in the passive YIG sphere. This dynamical process is accompanied by enhancement of the magneostrictive nonlinearity. As a result, an ultralow-threshold magnetic chaos is triggered in the broken $PT$ symmetry regime. Here the magnetic chaos with a prominent ultralow threshold is mainly benefited from the unconventional $PT$-symmetric system. This kind $PT$-symmetric structure may be applied in acoustic, electric and other fields helping to observe more valuable phenomena with the enhanced nonlinearities.
Parameters in the system are reachable referred to the latest experiments.

\begin{figure}[htbp]
\centering
\includegraphics[width=0.83\linewidth]{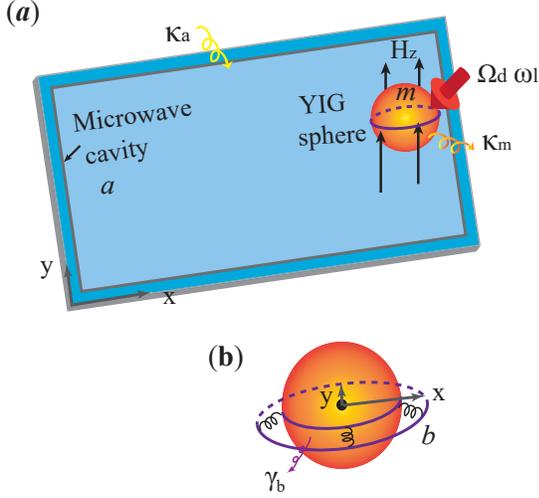}
\caption{(Color online) Schematic diagram of a $PT$-symmetry CMMS. (a) A three-dimensional active microwave cavity with a gain factor $\kappa_a$, where a millimetric YIG sphere is fixed in, is putted in an static magnetic field (not shown) along the $x$ direction. The YIG sphere is located in an uniform bias magnetic field $H_z$ (in the $z$ direction), which makes the magnon mode interacts with the active microwave cavity. A microwave pump field (not shown) is acted on the magnon mode to enhance the magnomechanical interaction. We drive the magnon mode directively with an strong microwave field $\Omega_d$ with frequency $\omega_d$. (b) the perspicuous diagrammatic drawing of the generation of mechanical vibration mode.}
\label{fig1}
\end{figure}

We consider a $PT$-symmetric CMMS, in which a Kittle-mode magnon mode (which is tuned at the resonant frequency $\omega_c=10.1\rm MHz$ with the cavity mode.) respectively couples to an active microwave cavity via magnetic dipole interaction and couples to a phonon mode by magnetostrictive interaction, as schematically shown in Fig\,\ref{fig1}.
The magnon resonates as excitation of collective spin in the YIG sphere (250-mm-diameter, i.e., Ref~\cite{Zhang2016}) placed in the uniform bias magnetic field. The uniform magnetic field ensures the cavity-magnon coupling. Meanwhile, due to the excellent material and geometrical features of the YIG sphere, the varied magnetization caused by the magnon excitation results in the spherical deformation of its geometric dimensioning, and generates mechanical vibration modes on the YIG sphere (vice versa)~\cite{Zhang2016}. We make an assumption that the microwave length is much larger than the size of the YIG to safely ignore the radiation pressure.
In a frame rotating with frequency $\omega_c$, the system Hamiltonian, with rotating wave approximation, given as
\begin{eqnarray}
\hat{H} &= &\frac{\hbar \omega_b}{2}({\hat{q}}^2+{\hat{p}}^2)+\hbar g_0 \hat{m}^\dag \hat{m} \hat{q}+\hbar g_{ma} ({\hat{a}}^\dag \hat{m}+\hat{a} \hat{m}^\dag)
\nonumber\\&&
+i\hbar \Omega_d (\hat{m}^\dag e^{-i\Delta t}-\hat{m}e^{i\Delta t}).
\label{e1}
\end{eqnarray}

Where $\hat{q}=\frac{1}{\sqrt{2}}(\hat{b}^\dag +\hat{b})$ and $\hat{p}=\frac{i}{\sqrt{2}}(\hat{b}^\dag -\hat{b})$ ($\hat{b}$ and $\hat{b}^\dag$ are the annihilation and creation operators of mechanical mode) stand for the position and momentum operators of the mechanical mode (with frequency $\omega_b$ ). The $\hat{a}$ ($\hat{a}^\dag$) and $\hat{m}$ ($\hat{m}^\dag$) denote the annihilation (creation) operators

\begin{figure}[htbp]
\centering
\includegraphics[width=0.95\linewidth]{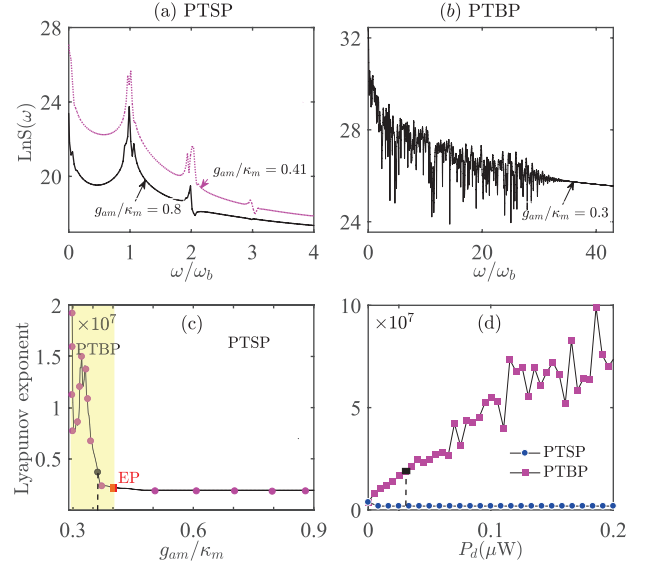}
\caption{(Color online) The spectrum ${\rm LnS}( \omega )$ of the magnon mode $I_m={m_r}^2+{m_i}^2$ (a) in $PT$SP, (b) in $PT$BP; and the insets are the trend of magnon-intensity variation in the time domain. The Lyapunov exponents versus (c) the magnon-cavity coupling rat $g_{ma}/\kappa_m$, (d) the microwave driving power $P_d/\kappa_m$; the insets present the spectrum of the magnon intensity at one point, respectively. We take the system parameters as $\kappa_m=2\times\pi 10^6$, $\kappa_a/\kappa_m=0.6$, $\gamma_b/\kappa_m=10^{-4}$, $g_0/\kappa_m=10^{-6}$, $\omega_b/\kappa_m=10$, and in (a)-(c) $P_d=0.21\rm m$W, in (d) $g_{am}/\kappa_m=0.2, 0.8$ for the pink and blue curves, respectively. The time interval in each subgraph is choose from 4$\mu$s-8$\mu$s.}
\label{fig2}
\end{figure}
of the active cavity mode and the lossy magnon mode, respectively. the magnomechanical coupling strength $g_0$ is usually very small and can be enhanced by a microwave driving (not shown) acting on the YIG~\cite{Zhang2016, Agarwal2018}. The photon-magnon coupling rate $g_{am}$ is in the strong regime~\cite{You2018}, because it is much larger than the dissipations $\kappa_a$, $\kappa_m$ of the microwave cavity and the magnon modes, respectively.
The magnetic frequency is tuned at will by the external bias magnetic $H_z$ and gyromagnetic ratio $\gamma$. The last term describes the unique monochrome driving (with amplitude $\Omega_d=\sqrt{\frac{\kappa_m P_d}{2\hbar \omega_d}}$) applied on the magnon mode. The frequency detuning $\Delta=\omega_d-\omega_c$ from the driving field and the magnon mode is resonant with the mechanical mode, this can fully excite the mechanical mode and reduces the driving power.

For simplicity of later calculation, the average value of the operator is defined as: $c=c_r+ic_i$ ($c$ represents operator of any cavity or magnon mode in the system, $c_r$ and $c_i$ are real numbers.).
With gain and loss process, the nonlinear dynamics of CMMS can be expressed as a nonlinear differential equation $\dot{\vec{u}}={\mathcal{A}}\vec{u}+{\mathcal{A}}_0$, with $\vec{u}= (q, p, m_r, m_i, a_r, a_i)$ and $A_0=(0, 0, \Omega_d {\rm cos}(-\Delta t), -\Omega_d  {\rm sin}(\Delta t), 0, 0)$. The coefficient matrix
\[{\mathcal{A}}=
\begin{pmatrix}
{0} & {\omega_b} & {0} & {0} & {0} & {0} \\
{-\omega_b} & {-\gamma_m} & {-\sqrt{2}g_0 m_r} & {-\sqrt{2}g_0 m_i} &  {0} &  {0} \\
{\frac{\sqrt{2}}{2}g_0 m_i} &  {0} & {\frac{\kappa_m}{2}} & {\frac{\sqrt{2}}{2}g_0 q} & {0} & {g_{am}} \\
-\frac{\sqrt{2}}{2}g_0 m_r & 0 & -\frac{\sqrt{2}}{2}g_0 q & -\frac{\kappa_m}{2} & -g_{am} & 0 \\
0 & 0 & 0 & g_{am} & \frac{\kappa_a}{2} & 0\\
0 & 0 & -g_{am} & 0 & 0 & \frac{\kappa_a}{2}
\end{pmatrix}.
\]

Where $\gamma_m$ is the loss rate of the phonon mode. From the second to the fourth row, the matrix elements including variates are the nonlinear terms derived from the magnomechanical interaction in the Hamiltonian. It illu
strates that
the mechanical deformation varies with the magnon intensity through the nonlinear interaction; inversely, the magnon mode is affected by the deformation. In this case, at the expense of a strong driving power, the magnon and mechanical modes can enter the chaotic regime. Nevertheless, in the present work, the magnomechanical nonlinearity can be dramatically amplified and the chaotic threshold can be decreased more than three orders of magnitude as shown in Fig.\,{ref{fig2}} (d).
Meanwhile the system phase is in the $PT$BP regime, the eigenfrequencies of the magnon and cavity modes after diagonalization become double degenerate frequency.

\begin{figure}[htbp]
\centering
\includegraphics[width=0.95\linewidth]{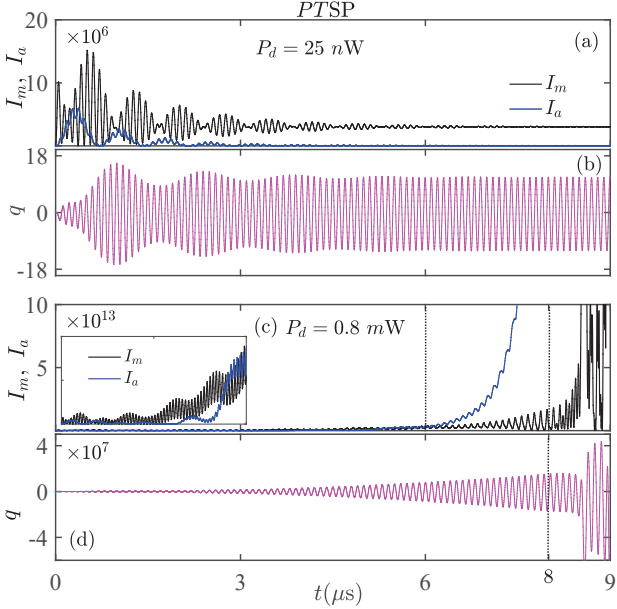}
\caption{(Color online) Evolution of (a) magnon intensity (lossy mode) and the cavity intensity (gain mode) and (b) the mechanical displacement with driving field $P_d=0.25~n$W in the $PT$SP. In (c)-(d) the difference from above is the driving strength $P_d=0.8~m$W, the inset clearly shows the evolution during the first six seconds.
The system parameters are the same as in Fig.\,\ref{fig2} except for $g_{am}/\kappa_m = 0.8$.}
\label{fig3}
\end{figure}

For sensitivity of chaos to the initial conditions, we explore the perturbation $\vec{\varepsilon} = ({\varepsilon}_{q}, {\varepsilon}_{p}, {\varepsilon}_{m_r}, {\varepsilon}_{m_i}, {\varepsilon}_{a_r}, {\varepsilon}_{a_i})$ defined by linearizing above nonlinear differential equation ($\dot{\vec{u}}$) to illustrate the chaos properties in CMMS. We get the linear differential equations $\dot{\vec{\varepsilon}}=\mathcal{L}\varepsilon$ and coefficient matrix $\mathcal{L}$ is wrote as
\[{\mathcal{L}}=
\begin{pmatrix}
{0} & {\omega_b} & {0} & {0} & {0} & {0} \\
{-\omega_b} & {-\gamma_m} & {-2\sqrt{2}g_0 m_r} & {-2\sqrt{2}g_0 m_i} &  {0} &  {0} \\
{\frac{\sqrt{2}}g_0 m_i} &  {0} & {\frac{\kappa_m}{2}} & {\frac{\sqrt{2}}g_0 q} & {0} & {g_{am}} \\
-\frac{\sqrt{2}}g_0 m_r & 0 & -\frac{\sqrt{2}}g_0 q & -\frac{\kappa_m}{2} & -g_{am}& 0 \\
0 & 0 & 0 & g_{am} & \frac{\kappa_a}{2} & 0\\
0 & 0 & - g_{am} & 0 & 0 & \frac{\kappa_a}{2}
\end{pmatrix}.
\]
Combining linear and nonlinear equations together, we can numerically calculate the Lyapunov exponent to judge the dynamical behavior of system.

By calculating the nonlinear differential equation, we analyze the intensity spectrum $I_b=m^2_r+m^2_i$ of the magnon mode in Fig.\,{\ref{fig2}} (a) in the $PT$SP regime and (b) in the $PT$BP regime with a driving power $P_d=0.21\,m$W. In subgraph (a) $g_{am}>\frac{\kappa_a+\kappa_m}{4}=0.4\kappa_m$ (equilibrium point), the spectrum is separated from each other with little branchings in the peak. This is resulted from the real part of eigenfrequencies $\sqrt{g^2_{am}-(\frac{\kappa_a+\kappa_m}{4})^2}$ after diagonalization, which also determines the width of the branchings depending on the values of $g_{am}$. In subgraph (b) $g_{am}<0.4\kappa_m$, the system falls into the
$PT$BP regime, the spectrum becomes continuous indicating the emergence of the chaos. The insets in (a) and (b) are the corresponding magnetic
intensities in time domain, which go through a period to chaos behaviors.

\begin{figure}[htbp]
\centering
\includegraphics[width=0.95\linewidth]{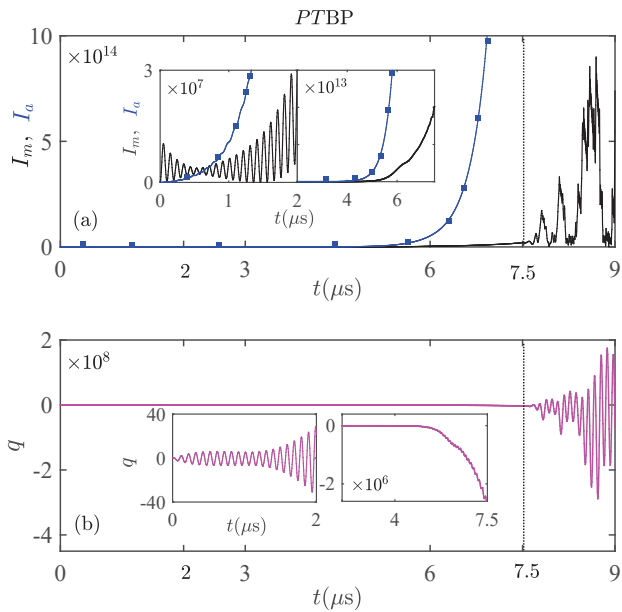}
\caption{(Color online)  Evolution of (a) magnon intensity (lossy mode) and the cavity intensity (gain mode) and (b) the mechanical displacement in the $PT$BP. The evolutionary processes before accessing the chaotic regime are amplified in the illustrations.
The system parameters are the same as in Fig.\,\ref{fig3} (a)-(b) except for $g_{am}/\kappa_m = 0.2$.}
\label{fig4}
\end{figure}
Here the driving field $P_d=0.21m$W is smaller at least two order of
magnitude than the chaotic threshold in the conventional CMMS. It is benefited from the localization-enhanced magnomechnaical nonlinearity when the phase transits from $PT$SP to $PT$BP.

Keep the driving strength $0.21\rm m$W, we demonstrate the influence of degree of the $PT$-symmetric breaking depending on the parameter $g_{am}/\kappa_m$ on the chaos by Lyapunov exponent. Lyapunov exponent is defined as the logarithmic slope of intensity of the perturbation $\varepsilon_{I_m}=|m+\varepsilon_m|^2-|m|^2$ as shown in Fig.\,{\ref{fig2}} (c). When $g_{am}/\kappa_m<0.4$ (equilibrium point: EP), the system enters into the $PT$BP and the Lyapunov exponent (through nonlinear enhancement) is in sharp contrast to that in $PT$BP.
The reduction of parameter $g_{am}/\kappa_m$ is accompanied by enhancement of degree of the $PT$-symmetric breaking, i.e., the localization-nonlinear-enhancement. So when $g_{am}/\kappa_m<0.3$, the Lyapunov exponent is too big to numerically calculate in Fig.\,{\ref{fig2}} (c).
In Fig.\,{\ref{fig1}} (d), Lyapunov exponent increasing with enhancement of the driving strength in $PT$SP with $g_{am}/\kappa_m=0.2$ is clearly distinguished from that in $PT$BP with $g_{am}/\kappa_m=0.8$. The spectrum of the inset in (d) corresponds to the black rectangle $P_d=25~n$W on the pink-dot curve. It means that we can get a ultra-low chaotic threshold with a deeper degree of $PT$-symmetric breaking.

In order to analysis the mechanism of ultra-low chaos threshold, we go into the evolutionary progresses of the intensity of magnon and cavity modes and the mechanical displacement (deformation size of the YIG sphere) over time in the $PT$SP and $PT$BP.

In $PT$SP, when the system is driven by a weak driving field
$P_d=25~n$W, the intensities $I_m$, $I_a$ and the mechanical displacement $q$ almost experience synchronous periodic oscillations in Fig.\,\ref{fig3} (a)-(b). This is because the magnon-loss is stronger than the magnetic tunneling in the weak driving regime, resulting in no cavity-localization. Fig.\,{\ref{fig3}} (c)-(d) show the case i.e., increasing the driving power to 0.8$~m$W on magnon mode. The inset magnifies the behavior during the first six seconds, $I_a$ of gain cavity mode oscillates periodically with rapid intensification, then exceeds $I_m$. Coming after that, it increases steeply with oscillation caused by the cavity-magnon interaction. With strong driving, the magnon mode is excited and periodically enhanced, which ultimately leads to the chaotic behavior of $I_m$ featuring the non-periodic oscillation. In harmony with $I_m$, mechanical displacement $q$ becomes chaotic after enhanced periodic oscillation. In $PT$SP, strong driving is needed to generate strong enough magnetostrictive nonlinearity for accessing chaotic regime of the system. Note that the driving threshold of the chaos will be much stronger in a conventional CMMS with a lossy cavity mode.

However, in $PT$BP, to access chaotic regime, strong driving is completely not needed replaced by adjusting the parameter $g_{am}$ representing the degree of $PT$-symmetric breaking. In Fig.\,{\ref{fig4}} we lock the weak driving strength $25~n$W and the parameter $g_{am}/\kappa_m=0.2$.
It demonstrates that the field-localization emerges via the exponential enhancement of the cavity energy $I_a$ (the blue curve). This results in an unidirectional energy transmission from cavity field to magnon mode, i.e., amounts of magnon is excited, whose intensity $I_m$ is pulled into exponential accumulation from periodical behavior. Large localized magnon in the YIG sphere induces enhanced magnomechnaical nonlinearity, lastly the chaotic behaviors appear both on the magnon and mechanical modes.


In conclusion, we mainly research the nonlinear magnomechanical dynamics in a $PT$-symmetric CMMS consisting of a gain microwave cavity and a lossy magnomechanical subsystem (YIG sphere). Through adjusting the $PT$ phase decided by the cavity-magnon interaction-to-loss rate, we generate the magnomechanical chaos featuring an ultra-low driving threshold. This feature is resulted from the localization-enhanced-nonlinear in the $PT$BP. Moreover, the chaos can be switched off and on by modulating the phase transition between $PT$SP and $PT$BP. Our work broadens the research scope of magnon and establish a versatile platform for information processing between photon, magnon and phonon. With currently experimental technology, our approach will be validated by experiments soon.




\begin{thebibliography}{0}%
\makeatletter
\providecommand \@ifxundefined [1]{%
 \@ifx{#1\undefined}
}%
\providecommand \@ifnum [1]{%
 \ifnum #1\expandafter \@firstoftwo
 \else \expandafter \@secondoftwo
 \fi
}%
\providecommand \@ifx [1]{%
 \ifx #1\expandafter \@firstoftwo
 \else \expandafter \@secondoftwo
 \fi
}%
\providecommand \natexlab [1]{#1}%
\providecommand \enquote  [1]{``#1''}%
\providecommand \bibnamefont  [1]{#1}%
\providecommand \bibfnamefont [1]{#1}%
\providecommand \citenamefont [1]{#1}%
\providecommand \href@noop [0]{\@secondoftwo}%
\providecommand \href [0]{\begingroup \@sanitize@url \@href}%
\providecommand \@href[1]{\@@startlink{#1}\@@href}%
\providecommand \@@href[1]{\endgroup#1\@@endlink}%
\providecommand \@sanitize@url [0]{\catcode `\\12\catcode `\$12\catcode
  `\&12\catcode `\#12\catcode `\^12\catcode `\_12\catcode `\%12\relax}%
\providecommand \@@startlink[1]{}%
\providecommand \@@endlink[0]{}%
\providecommand \url  [0]{\begingroup\@sanitize@url \@url }%
\providecommand \@url [1]{\endgroup\@href {#1}{\urlprefix }}%
\providecommand \urlprefix  [0]{URL }%
\providecommand \Eprint [0]{\href }%
\providecommand \doibase [0]{http://dx.doi.org/}%
\providecommand \selectlanguage [0]{\@gobble}%
\providecommand \bibinfo  [0]{\@secondoftwo}%
\providecommand \bibfield  [0]{\@secondoftwo}%
\providecommand \translation [1]{[#1]}%
\providecommand \BibitemOpen [0]{}%
\providecommand \bibitemStop [0]{}%
\providecommand \bibitemNoStop [0]{.\EOS\space}%
\providecommand \EOS [0]{\spacefactor3000\relax}%
\providecommand \BibitemShut  [1]{\csname bibitem#1\endcsname}%
\let\auto@bib@innerbib\@empty
\end{thebibliography}%


\begin{thebibliography}{99}
\bibitem{Chumak2015} A.~V. Chumak, V.~I. Vasyuchka, A.~A. Serga, and B.~Hillebrands,
``Magnon spintronics,''
Nat. Phys. \textbf{11}, 453 (2015).


\bibitem{Lenk2011} B.~Lenk, H.~Ulrichs, F.~Garbs, M. M{\"u}nzenberg,
``The building blocks of magnonics,''
Phys. Rep. \textbf{507}, 033819 (2005).

\bibitem{Osada2016} A.~Osada, R.~Hisatomi, A.~ Noguchi, Y.~Tabuchi, R.~Yamazaki, K.~Usami, M.~ Sadgrove, R.~Yalla, M.~Nomura, and Y.~Nakamura,
``Cavity Optomagnonics with Spin-Orbit Coupled Photons,''
Phys. Rev. Lett. \textbf{116}, 223601 (2016).

\bibitem{Cao2015} Y.~Cao, P.~Yan, H.~Huebl, S.~T.~ B. Goennenwein, and G.~E.~W. Bauer,
``Exchange magnon-polaritons in microwave cavities,''
Phys. Rev. B \textbf{91}, 094423 (2015).

\bibitem{Zhang2014} X.~Zhang, C.-L. Zou, L.~J. Hong, and X.~Tang,
``Strongly Coupled Magnons and Cavity Microwave Photons,''
Phys. Rev. Lett. \textbf{113}, 156401 (2014).

\bibitem{Yao2015} B.~M. Yao, Y.~S. Gui, Y.~Xiao, H.~Guo, X.~S. Chen, W.~Lu, C.~L. Chien, and C.-M. Hu,
``Theory and experiment on cavity magnon-polariton in the one-dimensional configurationm,''
Phys. Rev. B \textbf{92}, 184407 (2015).

\bibitem{Serga2010} A.~A. Serga, A.~V. Chumak, and B.~Hillebrands,
``YIG magnonics,''
J. Phys. D Appl. Phys. \textbf{43}, 264002 (2010).

\bibitem{Spencer1958} E.~G. Spencer, and R.~C. LeCraw,
``Magnetoacoustic Resonance in Yttrium Iron Garnet,''
Phys. Rev. Lett. \textbf{1}, 241 (1958).

\bibitem{You2018} Y.~P. Wang, G-Q. Zhang, D.~Zhang, T.-F. Li, C.-M. Hu, and J.~Q. You,
``Bistability of Cavity Magnon Polaritons,''
Phys. Rev. Lett. \textbf{120}, 057202 (2018);

\bibitem{Zhang2016} X.~Zhang, N.~Zhu, C.-L. Zou, and H.~X. Tang,
``Optomagnonic Whispering Gallery Microresonators,''
Phys. Rev. Lett. \textbf{117}, 123605 (2016);

\bibitem{Tang2016} X.~Zhang, C.-L. Zou, L.~Jiang, and H.~X. Tang,
``Cavity magnomechanics,''
Sci. Adv. \textbf{2}, 1501286 (2016);


\bibitem{Nakamura2014} Y.~Tabuchi, S.~Ishino, T.~Ishikawa, R.~Yamazaki, K.~Usami, and Y.~ Nakamura,
``Hybridizing Ferromagnetic Magnons and Microwave Photons in the Quantum Limit,''
Phys. Rev. Lett. \textbf{113}, 083603 (2014);

\bibitem{Agarwal2018} J.~Li, S.-Y. Zhu, and G.~S. Agarwal,
``Magnon-photon-phonon entanglement in cavity magnomechanics,''
arXiv:1807.07158 (2018);

\bibitem{Liu2018} Z.-X. Liu, B.~Wang, H.~Xiong, and Y. Wu,
``Magnon-induced high-order sideband generation,''
Opt. Lett. \textbf{43}, 3698 (2018);


\bibitem{Bender1998} C.~M. Bender, and S. Boettcher,
``Real Spectra in Non-Hermitian Hamiltonians Having $PT$ Symmetry,''
Phys. Rev. Lett. \textbf{80}, 5243 (1998);


\bibitem{Bender2007} C.~M. Bender,
vMaking sense of non-Hermitian Hamiltonians,''
Rep. Prog. Phys. \textbf{70}, 947 (2007);

\bibitem{Peschel2012} A. Regensburger, C.~Bersch, M.-A. Miri, G.~Onishchukov, D.~N. Christodoulides, and U. Peschel,
``Parity-time synthetic photonic lattices,''
Nature \textbf{488}, 167 (2012);

\bibitem{Peng2014} B.~Peng, {\c S}.~K. \"{O}zdemir, F.~Lei, F.~Monifi, M.~Gianfreda, G.~L. Long, S.~Fan,  F.~Nori, C.~M. Bender, and L.~Yang,
``Parity¨Ctime-symmetric whispering-gallery microcavities,''
Nat. Phys. \textbf{10}, 394 (2014);

\bibitem{Guo2009} A.~Guo, G.~J. Salamo, D.~Duchesne, R.~Morandotti, M.~Volatier-Ravat, V. ~Aimez, G.~A. Siviloglou, and D.~N. Christodoulides,
``Observation of $PT$-Symmetry Breaking in Complex Optical Potentials,''
Phys. Rev. Lett. \textbf{103}, 093902 (2014);

\bibitem{Yang2014} B.~Peng, {\c S}.~K. {\"O}zdemir, S.~Rotter, H.~Yilmaz, M.~Liertzer, F.~Monifi, C.~M. Bender, F.~Nori, and L.~Yang,
``Loss-induced suppression and revival of lasing,''
Science \textbf{346}, 328 (2014);

\bibitem{Feng2014} L.~Feng, Z.~J. Wong, R.-M. Ma, Y.~Wang, and X.~Zhang,
``Single-mode laser by parity-time symmetry breaking,''
Science \textbf{346}, 972 (2014);


\bibitem{Lv2015} X.-Y. L\"u, H.~Jing, J.-Y. Ma, and Y.~Wu,
``$PT$-Symmetry-Breaking Chaos in Optomechanics,''
Phys. Rev. Lett. \textbf{114}, 253601 (2015).

\bibitem{Yang2017} W.~Chen, {\c S}.~K. {\"o}zdemir, G.~Zhao, J.~Wiersig, and L.~Yang,
``Exceptional points enhance sensing in an optical microcavity,''
Nature \textbf{548}, 192 (2017).


\bibitem{Hodaei2017} H.~Hodaei, A.~U. Hassan, S.~Wittek, H.~Garcia-Gracia, R.~El-Ganainy, D.~N. Christodoulides, and M.~ Khajavikhan,
``Enhanced sensitivity at higher-order exceptional points,''
Nature \textbf{548}, 187 (2017).

\bibitem{You2017} D.~Zhang, X.-Q. Luo, Y.-P. Wang, T.-F. Li, and J.~Q. You,
``Observation of the exceptional point in cavity magnon-polaritons,''
Nat. Commun. \textbf{8}, 1368 (2017).

\end{thebibliography}

\end{document}